\documentclass{IEEEtran}
\usepackage{amsmath,amsfonts}
\usepackage{mathtools}
\usepackage{algorithmic}
\usepackage{algorithm}
\usepackage{array}
\usepackage[caption=false,font=normalsize,labelfont=sf,textfont=sf]{subfig}
\usepackage{textcomp}
\usepackage{anyfontsize}
\usepackage{stfloats}
\usepackage{hyperref}
\usepackage{graphicx}
\usepackage{cite}
\usepackage{bm}
\usepackage{nomencl}
\usepackage{multirow}
\usepackage{booktabs}
\usepackage{tikz}
\usepackage{makecell}

\newif\ifmarkup
\markupfalse %
\ifmarkup
	\usepackage[authormarkup=none]{changes}
	\usepackage[normalem]{ulem} %
	\definechangesauthor[name={new}]{new}
	\definechangesauthor[name={deletion}]{deletion}
	\definechangesauthor[name={modification}]{modification}
	\newcommand\xadded{\added[id=new]}
	\newcommand\xdeleted{\deleted[id=deletion]}
	\newcommand\xreplaced{\replaced[id=modification]}
	\setaddedmarkup{\textcolor{blue}{#1}}
	\setdeletedmarkup{\textcolor{red}{\sout{#1}}}
	\makeatletter
	\patchcmd{\Changes@output}{%
	\IfStrEq{#1}{replaced}{{\Changes@Markup@added{#3}}\allowbreak\Changes@Markup@deleted{#4}}{}%
	}{%
	\IfStrEq{#1}{replaced}{{\Changes@Markup@deleted{#4}}\allowbreak\Changes@Markup@added{#3}}{}%
	}{}{}
	\makeatother
\else
	\usepackage[final]{changes}
	\newcommand\xadded{\added}
	\newcommand\xdeleted{\deleted}
	\newcommand\xreplaced{\replaced}
\fi

\usetikzlibrary{shapes.geometric, arrows.meta, positioning, fit}

\tikzstyle{startstop} = [rectangle, rounded corners, minimum width=3cm, minimum height=1cm,text centered, draw=black, fill=gray!20]
\tikzstyle{process} = [rectangle, minimum width=3cm, minimum height=1cm, text centered, draw=black, fill=blue!10]
\tikzstyle{arrow} = [thick, ->, >=Stealth]
\def\BibTeX{{\rm B\kern-.05em{\sc i\kern-.025em b}\kern-.08em
		T\kern-.1667em\lower.7ex\hbox{E}\kern-.125emX}}
\newcommand{\balpha}{\bm{\alpha}}
\newcommand{\bmu}{\bm{\mu}}
\newcommand{\bSigma}{\bm{\Sigma}}

\newcommand{\bC}{\bm{C}}

\newcommand{\bd}{\bm{d}}
\newcommand{\ba}{\bm{a}}

\newcommand{\bp}{\bm{p}}
\newcommand{\bx}{\bm{x}}

\newcommand{\bz}{\bm{z}}
\newcommand{\bA}{\bm{A}}
\newcommand{\bone}{\bm{1}}
\newcommand{\bH}{\bm{H}}
\newcommand{\bh}{\bm{h}}
\newcommand{\bxi}{\bm{\xi}}

\newcommand{\boldeta}{\bm{\eta}}
\newcommand{\bnu}{\bm{\nu}}
\newcommand\C[1]\null
\begin{document}
	
	\title{Chance-Constrained DC Optimal Power Flow Using Constraint-Informed Statistical Estimation}
	
	\author{Tianyang Yi,~\IEEEmembership{Student Member,~IEEE}, D. Adrian Maldonado,~\IEEEmembership{Member,~IEEE}, Anirudh Subramanyam
		\thanks{This material is based upon work supported by the National Science Foundation under Grant No. DMS-2229408 and the U.S. Department of Energy, Office of
			Science, under Contract DE-AC02-06CH11357.}
		\thanks{Tianyang Yi and Anirudh Subramanyam are with The Pennsylvania State University, University Park, PA, USA. (email: \url{subramanyam@psu.edu}).}
		\thanks{D. Adrian Maldonado is with Argonne National Laboratory, Lemont, IL, USA, and The University of Chicago, Chicago, IL, USA.}}

	\maketitle

	\begin{abstract}
		Chance-constrained optimization has emerged as a promising framework for managing uncertainties in power systems. This work advances its application to the DC Optimal Power Flow (DC-OPF) model, developing a novel approach to uncertainty modeling and estimation. Current methods typically tackle these problems by first modeling random nodal injections using high-dimensional statistical distributions that scale with the number of buses, followed by deriving deterministic reformulations of the probabilistic constraints. We propose an alternative methodology that exploits the constraint structure to inform the uncertainties to be estimated, enabling significant dimensionality reduction. Rather than learning joint distributions of net-load forecast errors across units, we instead directly model the one-dimensional aggregate system forecast error and two-dimensional line errors weighted by power transfer distribution factors. We evaluate our approach under both Gaussian and non-Gaussian distributions on synthetic and real-world datasets, demonstrating significant improvements in statistical accuracy and optimization performance compared to existing methods.
	\end{abstract}
	
	\begin{IEEEkeywords}
		Chance Constraints, DC Optimal Power Flow
	\end{IEEEkeywords}
	
	\section*{Nomenclature}
	\addcontentsline{toc}{section}{Nomenclature}
	\begin{IEEEdescription}[\IEEEusemathlabelsep\IEEEsetlabelwidth{$\hat{\lambda}_k, \hat{\bnu}_k, \hat{\bC}_k$}]%
		\item [$\mathcal{B}$] set of buses
		\item [$\mathcal{L}$] set of lines
		\item [$\mathcal{G}$] set of controllable generators
		\item [$\mathcal{D}$] set of loads (demands)
		\item [$\mathcal{G}_i$] set of generators ($\subseteq \mathcal{G}$) located at bus $i\in\mathcal{B}$
		\item [$\bH$] $|\mathcal{L}|\times|\mathcal{B}|$ power transfer distribution factors %
		\item [$\bh^{\text{wind}}_l, \bh^{\text{gen}}_l$] $|\mathcal{B}|$- and $|\mathcal{G}|$-dim vectors of power transfer distribution factors mapping wind and controllable unit power outputs to flow along line $l$
		\item[$c_{1,g}$,$c_{2,g}$] linear/quadratic cost coefficient of generator $g$
		\item[$\bar{p}^{\text{wind}}_i$] total wind power forecast at bus $i$
		\item [$\xi_i$] wind power forecast error at bus $i$
		\item[$\bar{d}_i$] load forecast at bus $i$
		\item[$p_g^{\text{min}}$,$p_g^{\text{max}}$] min/max output of generator $g$
		\item[$f_l^{\text{max}}$] capacity of line $l$
		\item[$\alpha_g$] participation factor of generator $g$
		\item[$\bar{p}_g, p_g(\bxi)$] scheduled and actual output of generator $g$
		\item[$p_i^0,\Delta p_i(\bxi)$] nominal and deviation of injection at bus $i$
		\item[$f_l^0,\Delta f_l(\bxi)$] nominal and deviation of flow along line $l$
		\item[$p_i(\bxi)$] actual net power injection into bus $i$
		\item[$f_l(\bxi)$] actual flow along line $l$
		\item[$\gamma_l(\balpha)$] flow deviation along $l$ from generator control
		\item[$\Lambda_l$] flow deviation along $l$ from forecast errors
		\item [$\Omega$] random variable of total system forecast error
		\item [$\boldeta_l$] random two-dim vector affecting line $l$ flow %
		\item[$\hat{\pi}_k, \hat{\bmu}_k, \hat{\bSigma}_k$] Gaussian mixture parameter estimates of $\bxi$
		\item[$\hat{\beta}_k, \hat{m}_k, \hat{\sigma}_k^2$] Gaussian mixture parameter estimates of $\Omega$
		\item[$\hat{\lambda}_k, \hat{\bnu}_k, \hat{\bC}_k$] Gaussian mixture parameter estimates of $\boldeta_l$
		\item [$\epsilon$] maximum acceptable violation probability %
		\item [$\Phi$] standard normal distribution function
		\item [$\hat{\Phi}, a_s, b_s$] piecewise linear approximation of $\Phi$ and slope and intercept of the $s^\text{th}$ linear segment
	\end{IEEEdescription}
	
	\section{Introduction}
	Power system planning has become increasingly complex due to the presence of uncertain energy sources.
	Uncertainties arise from variability and unpredictability in the system,
	specifically due to forecast errors in net load and generation. 
	These forecast errors directly affect the feasibility of deterministic planning procedures, such as those resulting from Optimal Power Flow (OPF) analysis, and must be accounted for\cite{xie2010wind}.
	
	To address these challenges, several approaches have been developed, including robust optimization \cite{sousa2010robust,louca2018robust}, multi-stage stochastic programming \cite{chen2005multi}, and chance-constrained optimization \cite{zhang2011chance, bienstock2014chance}. Chance-constrained optimization incorporates constraints that enforce certain decision-dependent random events, such as line overloads, to occur with probability no greater than a prespecified risk level. Within the OPF framework, this approach allows system operators to specify risk tolerances for power generation or transmission line violations. %

	Chance-constrained optimization problems are typically solved using one of two broad classes of methods: sample-based \cite{lukashevich2023importance,calafiore2006scenario} and sample-free analytical methods \cite{shapiro2021lectures}. The former rely on generating a finite set of samples that represent possible realizations of the uncertain parameters; this allows converting the intractable probabilistic constraints into a deterministic form that can be solved by standard solvers. However, their key drawback is that a large number of samples may be required to obtain accurate solutions, making them computationally prohibitive for large-scale systems \cite{henrion2013:sample_size,sakhavand2020new}. On the other hand, sample-free methods reformulate the chance constraints into deterministic form using analytical approximations to the original problem, usually by assuming that the randomness follows certain probability distributions, such as Gaussian.
	
	In fact, existing chance-constrained OPF (CC-OPF) models typically assume a Gaussian distribution for net load forecasting errors \cite{bienstock2014chance, roald2017chance,lubin2019chance}. From a statistical perspective, the Gaussian assumption is appealing because the maximum likelihood estimation (MLE) of its parameters (i.e., mean and covariance) admits closed-form solutions, making parameter estimation computationally straightforward \cite{kay1993fundamentals}.
	Additionally, from an optimization perspective, assuming Gaussian uncertainties enables tractable convex reformulations of the chance constraints as second-order cone programs (SOCPs), which can be efficiently solved with modern optimization solvers \cite{bienstock2014chance}. 
	
	However, several statistical analyses have shown that non-Gaussian distributions, such as Weibull \cite{genc2005estimation}, Cauchy \cite{hodge2011wind}, \xadded{generalized Pareto \cite{mararakanye2022characterizing},} (and Cauchy-like non-Gaussian distribution errors \cite{vrakopoulou2013probabilistic}, \cite{zhang2024efficient}), may be better capable of representing wind generation power forecast errors, particularly within the shorter time horizon of OPF applications. Unfortunately, from the optimization perspective, the incorporation of non-Gaussian distributions is not as computationally straightforward as in the Gaussian case, since the resulting CC-OPF problems do not admit tractable reformulations that can be solved by existing solvers.
	
	A promising path for modeling these non-Gaussian forecast errors is via Gaussian Mixture Models (GMMs), as they can approximate a wide class of probability distributions by mixing a sufficiently large number of Gaussian components \cite{mclachlan2019finite}.
	At the same time, the analytical reformulation technique for Gaussian chance constraints has also been recently extended to the GMM setting \cite{wang2017chance, cui2017statistical, ke2015novel, valverde2012stochastic, fathabad2023asymptotically,song2024chance}, enabling solution of these models using standard solvers. 
	\xadded{The key advantage of GMMs is that linear combinations of GMM-distributed random variables remain GMM-distributed, allowing chance constraints to be reformulated analytically using component-specific Gaussian parameters. In contrast, non-GMM distributions such as Weibull, Cauchy, or generalized Pareto, typically do not preserve their distributional form under linear combinations (e.g., the sum of two Weibull random variables is not Weibull), precluding direct analytical reformulation. Therefore, we use GMMs to approximate forecast errors, enabling both analytical reformulations and the ability to capture non-Gaussian behavior.}

	\xadded{In this vein, \cite{wang2017chance} fits a GMM to the joint probability density of wind power output across multiple farms for chance-constrained economic dispatch; \cite{ke2015novel} uses a customized GMM to approximate the probability density of wind power generation for probabilistic OPF; \cite{valverde2012stochastic} uses GMMs as input distributions in stochastic power flow and state estimation of distribution networks; \cite{cui2017statistical} uses a generalized GMM to characterize wind power ramp distributions; and \cite{fathabad2023asymptotically} and \cite{song2024chance} fit GMMs to wind power generation at individual units for CC-OPF reformulation. In all these approaches, GMMs are fitted to the raw uncertainty at the individual component level, and the resulting distribution parameters are subsequently transformed to evaluate and reformulate the chance constraints.}

	\xreplaced{However, a key drawback of this paradigm is that the joint distribution of component-level uncertainties is high-dimensional.}{However, a key drawback of these and other existing approaches is that they use a \emph{multi-dimensional} GMM to model forecasting errors.} In particular, the dimension of the forecasting error distribution scales with the number of buses with stochastic injections (e.g., wind farms). From a statistical standpoint, this approach presents several challenges. First, standard MLE algorithms become significantly harder for high-dimensional GMMs. Unlike in the case of Gaussians, the MLE objective function is now nonconvex, and the Expectation-Maximization (EM) algorithm typically used in this context is only heuristic and prone to converge to local solutions \cite{bishop2006pattern}.
	Second, the problem is affected by the curse of dimensionality: the number of parameters to be estimated can grow so large that historical data becomes insufficient, %
	resulting in models that are prone to overfitting and that fail to generalize to unseen future data \cite{mclachlan2000finite,donoho2000high}. %
	In particular, roughly each additional $n$-dimensional Gaussian component in a GMM requires $O(n^2)$ new parameters to be estimated.
	
	Although dimensionality reduction techniques such as principal component analysis \cite{abdi2010principal} or latent variable models can reduce this computational burden, they are problem-structure-agnostic and do not account for how the uncertainties actually propagate through the OPF constraints. As a result, these methods risk discarding spatial correlations among wind forecast errors or underestimating extreme events, potentially leading to constraint violations and unreliable operations. There is a need to develop an integrated framework that aligns statistical modeling with the structure of the optimization constraints and out-of-sample risk of solutions.
	
	To address this gap, we propose a \emph{constraint-informed} approach, where the OPF constraints are used to identify \emph{low-dimensional} uncertainty structures that directly impact constraint feasibility. %
	Our work highlights the importance of statistical fitting in CC-OPF problems. Instead of fitting a joint high-dimensional distribution of net-load forecast errors at system buses, we use the structure of the chance constraints to identify one- and two-dimensional uncertainties that directly impact constraint feasibility. This reduces statistical complexity %
	and alleviates the difficulties associated with high-dimensional model fitting. Notably, the improvement stems not from modifying existing statistical estimation algorithms, but from identifying uncertainty representations relevant to the CC-OPF constraints. We demonstrate through synthetic and real-world forecast error data that this procedure improves estimation accuracy. We also illustrate the limitations of using the EM algorithm for statistical fitting, and how one can achieve better fits by modifying it from an OPF viewpoint. 
	
	Building on the constraint-informed framework, we also develop tractable analytical reformulations of chance constraints when the low-dimensional uncertainties follow a GMM. Through case studies, we show that it is the combination of constraint-informed dimensionality reduction and Gaussian mixture modeling that delivers significantly improved out-of-sample reliability compared to existing \xreplaced{approaches}{approahces}. We find that the improvement is particularly stark when the underlying forecast errors exhibit unimodal but heavy-tailed behavior, elucidating that GMMs %
	can also reduce out-of-sample risk in systems that may not involve multimodal data.
	
	The rest of the paper proceeds as follows. Section~\ref{sec:math_model} presents the model and assumptions; Section~\ref{sec:reformulation} presents the constraint-informed methodology, and provides reformulation of the chance constraints under Gaussian and GMM distributions; Section~\ref{sec:numerical} presents results of numerical experiments %
	on synthetic and real-world datasets. Throughout, we use boldface letters for vectors and matrices, and normal font for scalars.
	
	\section{Chance-Constrained DC-OPF Model}\label{sec:math_model}
	We consider a CC-OPF model with DC approximations that minimizes total generation costs while ensuring transmission line and generation limits are satisfied with high probability. 
	
	For simplicity, we assume that the demand vector $\bar{\bd}$ is not random and that it can be estimated with high accuracy. This is because demand fluctuations often occur on a different timescale than the decision-making window for OPF (e.g., 5--15 minutes). To model uncertainty, we associate each wind unit with a continuous random variable representing its forecast error. These errors are then collected into a random vector $\bxi\in\mathbb{R}^{|\mathcal{B}|}$ so that $\xi_i$ represents the total wind power forecast error at bus $i \in \mathcal B$. The actual power output at bus $i$ is thus $\bar{p}^{\text{wind}}_i+\xi_i$, where $\bar{p}_i^{\text{wind}}$ is the forecast production.
	Note that we set $\bar{p}^{\text{wind}}_i = \xi_i = 0$ if there are no wind units located at bus $i$.
	The (one-dimensional) aggregate system forecast error is:
	\begin{equation}
		\Omega\coloneqq\bone^\top \bxi=\sum_{i\in\mathcal{B}}\xi_i. 
		\label{eq:1d_rand} 
	\end{equation}
	
	The system operator must determine the nominal power dispatch of controllable generators, denoted by $\bar{p}_g$, under perfect forecasts (i.e., $\bxi=\bm{0}$). These dispatch levels will serve as baseline generation and they must satisfy the expected power balance. When the actual wind power outputs realize in real time, the controllable generators will adjust their output to compensate for the resulting power imbalance. These adjustments are made under the Automatic Generation Control (AGC) policy \cite{borkowska1974probabilistic}, where each generator contributes proportionally to the aggregate system-wide forecast error $\Omega$.
	In line with existing literature \cite{bienstock2014chance}, \cite{pena2020dc}, we model this AGC policy by assigning a participation factor $\alpha_g\geq 0$ to each controllable generator $g\in\mathcal{G}$. Under this policy, the actual power output of generator $g$ is thus stochastic and equal to:
	\begin{equation}
		p_g(\bxi)\coloneqq\bar{p}_g - \alpha_g\Omega.
		\label{eq: actual_power}
	\end{equation}
	When $\sum_{g \in \mathcal{G}} \alpha_g = 1$, equation~\eqref{eq: actual_power} ensures that the total adjustment matches system-wide imbalance. Furthermore, using~\eqref{eq: actual_power}, the net power injection into the system at bus $i\in\mathcal{B}$ is:
	\begin{equation}
		p_i(\bxi) = \sum_{g \in \mathcal{G}_i} (\bar{p}_g - \alpha_g\Omega) + (\bar{p}^{\text{wind}}_i+\xi_i) - \bar{d}_i.
		\label{eq:net_power_inj}
	\end{equation}
	We decompose this expression into a deterministic component and a stochastic deviation. First, we define the nominal power injection at bus $i$, under perfect forecast ($\bxi=\bm{0}$), as
	\begin{equation}
		p_i^0\coloneqq\sum_{g \in \mathcal{G}_i} \bar{p}_g +  \bar{p}^{\text{wind}}_i - \bar{d}_i.
		\label{eq: p_i^0}
	\end{equation}
	We can then model the stochastic deviation as
	\begin{equation}
		\Delta p_i(\bxi) \coloneqq -\sum_{g \in \mathcal{G}_i} \alpha_g\Omega + \xi_i.
		\label{eq: delta_power}
	\end{equation}
	This captures two effects happening at bus~$i$ due to uncertainty. The first term reflects \emph{global} balancing response from generators at bus $i$ to offset the aggregate system-wide forecast error $\Omega$, and the second term is simply the \emph{local} wind forecast error at bus~$i$. Using this notation, the net power injection under uncertainty can then be compactly expressed as
	\begin{equation}
		p_i(\bxi)=p_i^0 + \Delta p_i(\bxi).
		\label{eq: compact-power}
	\end{equation}
	
	The total power flow across line~$l$ can then be expressed as a linear function of the net power injections using the matrix of Power Transfer Distribution Factors (PTDF) \cite{wood2013power}:
	\begin{equation}
		f_l(\bxi) %
		=\sum_{i\in\mathcal{B}}H_{li}p_i(\bxi).
		\label{eq:flow} 
	\end{equation}
	As before, we can use \eqref{eq: p_i^0} and \eqref{eq: delta_power}, to define the nominal flow along line $l$ under perfect forecast conditions as:
	\begin{equation}
		f_l^0\coloneqq\sum_{i\in\mathcal{B}}H_{li}p_i^0.
		\label{eq:nominal_flow}
	\end{equation}
	The deviation of power injection $\Delta p_i(\bxi)$ is also propagated linearly into transmission lines via the PTDF matrix. We define the resulting stochastic deviation in line flow as:
	\begin{equation}
		\Delta f_l(\bxi)\coloneqq\sum_{i\in\mathcal{B}}H_{li}\Delta p_i (\bxi). 
		\label{eq:delta_flow}
	\end{equation}
	This stochastic deviation measures how the line flows are affected by forecast errors, both globally and locally. Substituting \eqref{eq: delta_power} into \eqref{eq:delta_flow} yields:
	\begin{equation}
		\Delta f_l(\bxi)=\bigg(-\sum_{i\in \mathcal{B}}H_{li}\sum_{g\in\mathcal{G}_i}\alpha_g\bigg)\Omega + \sum_{i\in \mathcal{B}}H_{li}\xi_i.
		\label{eq:del_flow_rewrite}
	\end{equation}
	In this expression, the coefficient of $\Omega$ represents the line flow deviation induced by global automatic control response from all controllable generators. We denote it as:
	\begin{equation}
		\gamma_l(\balpha)\coloneqq-\sum_{i\in \mathcal{B}}H_{li}\sum_{g\in\mathcal{G}_i}\alpha_g = -(\bh^{\text{gen}}_l)^\top \balpha.
		\label{eq: gammaa_l}
	\end{equation}
	The second component represents the line flow deviation induced by local wind forecast errors projected onto line $l$ via the PTDF matrix. We denote it as:
	\begin{equation}
		\Lambda_l\coloneqq\sum_{i\in \mathcal{B}}H_{li}\xi_i = (\bh^{\text{wind}}_l)^\top \bxi.
		\label{eq:Lambda_l}
	\end{equation}
	These terms allow us to express the total flow along line $l$ as a linear function of global and local uncertainty components:
	\begin{equation}
		f_l(\bxi) = f_l^0 + \Delta f_l(\bxi) = f_l^0 + \begin{bmatrix}
			\gamma_l(\balpha) & 1 
		\end{bmatrix}\begin{bmatrix}
			\Omega \\
			\Lambda_l 
		\end{bmatrix}.
		\label{eq:flow_matrix}
	\end{equation}
	Note that $\gamma_l(\balpha)$ is a deterministic decision-dependent constant. Therefore, the only source of randomness in \eqref{eq:flow_matrix} is:
	\begin{equation}
		\boldeta_l \coloneqq (\Omega, \, \Lambda_l) = (
		\bone^\top \bxi, \, (\bh^{\text{wind}}_l)^\top \bxi) \in \mathbb R^2.
		\label{eq: boldeta_l}
	\end{equation}
	The one- and two-dimensional random variables, $\Omega$ and $\boldeta_l$, will be the focus in the methodology section.
	
	We note that the use of the PTDF matrix implicitly ensures power balance at the bus level. Therefore, our model follows existing literature (e.g., see~\cite{song2024chance,pena2020dc}) and avoids the need to explicitly include phase angles as decision variables. Indeed, the resulting model only has the nominal power generation $\bar{\bp}$ and participation factors $\balpha$ as decision variables. Once their optimal values are obtained, the voltage phase angles (which are also random variables) can be readily recovered under a specific realization of $\bxi$ by first constructing $\bp(\bxi)$ using~\eqref{eq:net_power_inj} and then using the system admittance matrix. %
	
	We can now formulate the CC-OPF problem as follows:
	\begin{subequations}\label{eq:cc_opf}
		\begin{align}
			\min_{\bar{\bp}, \balpha} \quad & \mathbb{E}_{\bxi}\Bigg[\sum_{g\in\mathcal{G}}c_{2,g}p_g(\bxi)^2+c_{1,g}p_g(\bxi)\Bigg]
			\label{eq:cc_opf_obj} \\
			\text{s.t.} \quad 
			& \sum_{g \in \mathcal{G}} \alpha_g = 1, \quad \balpha \geq 0, \quad \bar{\bp} \geq 0,
			\label{eq:cc_opf_affine}\\
			& \sum_{g \in \mathcal{G}} \bar{p}_g + \sum_{i \in \mathcal{B}} \bar{p}^{\text{wind}}_i - \sum_{i \in \mathcal{B}} \bar{d}_i = 0,
			\label{eq:cc_opf_power_balance}\\
			& \mathbb{P} \left( p_g(\bxi)
			\geq p_g^{\min}  \right) \geq 1 - \epsilon, & \forall g \in \mathcal{G},
			\label{eq:cc_opf_power_min}\\
			&\mathbb{P} \left( p_g(\bxi)
			\leq p_g^{\max}  \right) \geq 1 - \epsilon,  & \forall g \in \mathcal{G},
			\label{eq:cc_opf_power_max}\\
			& \mathbb{P} \left(  f_l(\bxi) \geq -f_{l}^{\max} \right) \geq 1 - \epsilon, & \forall l \in \mathcal{L},
			\label{eq:cc_opf_line_flow_min}\\
			&\mathbb{P} \left( f_l(\bxi) \leq f_{l}^{\max}   \right) \geq 1 - \epsilon, & \forall l \in \mathcal{L}.
			\label{eq:cc_opf_line_flow_max}
		\end{align}
	\end{subequations} 
	The objective function \eqref{eq:cc_opf_obj} minimizes the expected total generation costs. It can be readily shown that it simplifies to:
	\begin{equation}
		\sum_{g \in \mathcal{G}}c_{2,g} \left(\bar{p}_g - \alpha_g\mathbb{E}[\Omega]\right)^2 + c_{2,g} \alpha_g^2 \mathbb{V}[\Omega] + c_{1,g} ( \bar{p}_g - \alpha_g\mathbb{E}[\Omega]).
		\label{eq: obj_final}
	\end{equation}
	Note that if we assume $\xi_i$ are i.i.d. zero-mean Gaussian random variables with known variance, then \eqref{eq: obj_final} further simplifies and reduces to the objective function used in prior work \cite{bienstock2014chance,roald2017chance}. %
	Constraints~\eqref{eq:cc_opf_affine} and~\eqref{eq:cc_opf_power_balance} together ensure power balance within the network under the affine control policy. Constraints~\eqref{eq:cc_opf_power_min} and~\eqref{eq:cc_opf_power_max} are chance constraints enforcing that the power generated lies within the minimum and maximum generation limits with (sufficiently high) probability $1-\epsilon$. Similarly, constraints~\eqref{eq:cc_opf_line_flow_min} and~\eqref{eq:cc_opf_line_flow_max} are chance constraints enforcing power flows along lines to remain less than the maximum line capacity with probability $1-\epsilon$. Here, $\epsilon$ is a risk parameter to be determined by the system operator. Lower values of $\epsilon$ reflect higher risk aversion, since they enforce a higher probability of constraint satisfaction. \xadded{In practice, different $\epsilon$ values may be assigned to different classes of constraints to reflect their operational importance.} We refer the reader to related works~\cite{lubin2015robust,roald2017chance,dvorkin2019chance,lubin2019chance} for \xadded{further discussion on} setting $\epsilon$ values\xadded{; in our experiments, we use a common value across all constraints so that the comparison focuses on the estimation methodology}.

	\section{Constraint-Informed Reformulation}\label{sec:reformulation}

    \xadded{Let $\bz = (\bar{\bp}, \balpha)$ denote the decision variables of~\eqref{eq:cc_opf}. Each chance constraint in~\eqref{eq:cc_opf_power_min}--\eqref{eq:cc_opf_line_flow_max} can be written in the form
    \begin{equation}
        \mathbb{P}\big(\ba(\bz)^\top \bA \bxi \leq b(\bz)\big) \geq 1-\epsilon,
        \label{eq:cc_generic_form}
    \end{equation}
    where $\ba(\bz)$ and $b(\bz)$ are affine functions of $\bz$ and $\bA$ is a constant matrix. For the power generation constraints~\eqref{eq:cc_opf_power_max}--\eqref{eq:cc_opf_power_min}, this follows from~\eqref{eq:1d_rand} and~\eqref{eq: actual_power} with $\bA = \bone^\top \in \mathbb{R}^{1 \times |\mathcal{B}|}$, so that $\bA\bxi = \Omega \in \mathbb{R}$. For the line flow constraints~\eqref{eq:cc_opf_line_flow_max}--\eqref{eq:cc_opf_line_flow_min}, this follows from~\eqref{eq:nominal_flow}, \eqref{eq:flow_matrix}, and~\eqref{eq: boldeta_l} with
	$
        \bA = \begin{bmatrix} \bone & \bh_l^{\mathrm{wind}} \end{bmatrix}^\top \in \mathbb{R}^{2 \times |\mathcal{B}|},
	$
    so that $\bA\bxi = \boldeta_l \in \mathbb{R}^2$.}
	
	\xadded{The key observation is that the constraint~\eqref{eq:cc_generic_form} depends on uncertainty only through the projection $\bA\bxi$, which is a low-dimensional linear function of the raw uncertainty $\bxi$. To evaluate the probability, one only needs the distribution of $\bA\bxi$ and not of $\bxi$ itself. Since $\bA\bxi$ is one-dimensional for power generation and two-dimensional for line flow constraints, learning its distribution is significantly more efficient and accurate than learning the full joint distribution of $\bxi \in \mathbb{R}^{|\mathcal{B}|}$.}

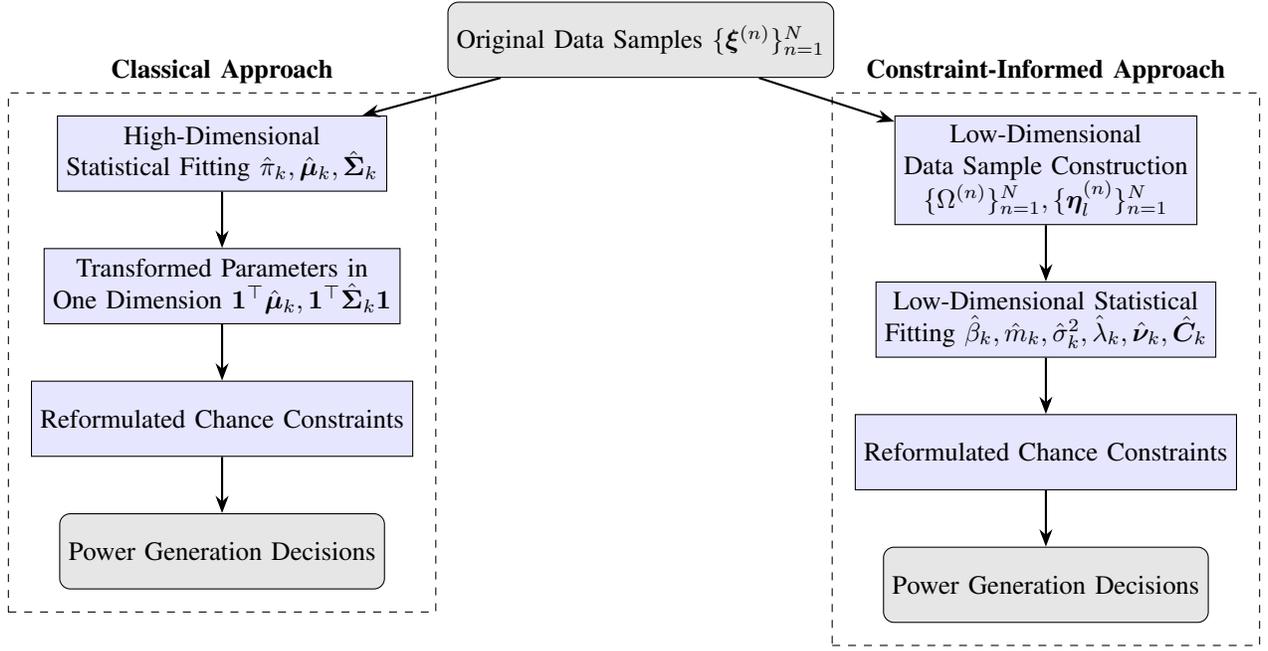
\begin{figure*}[t]
		\centering
		\begin{tikzpicture}[node distance=0.75cm and 0.75cm]
			
			\node (start) [startstop, align=center] %
			{Raw Forecast Error Data $\{\bxi^{(n)}\}_{n=1}^N$};
			
			\node (c1) [process, below left=0.5cm and 0.8cm of start, align=center] %
			{Fit high-dimensional Gaussian \\ mixture distribution to $\bxi$};
			\node (c2) [process, below=of c1, align=center] %
			{Transform fitted parameters \\ to obtain distribution of $\bA\bxi$};
			\node (c3) [process, below=of c2] {Reformulated Chance Constraints};
			\node (c4) [startstop, below=of c3] %
			{Optimal Dispatch Decisions};

			\draw [arrow] (start) -- node[midway, above left, xshift=-1pt] {} (c1);
			\draw [arrow] (c1) -- node[midway, left] {} (c2);
			\draw [arrow] (c2) -- (c3);
			\draw [arrow] (c3) -- (c4);

			\node (ci1) [process, below right=0.5cm and 0.8cm of start, align=center] %
			{Transform data samples: \\ $\{\bA\bxi^{(n)}\}_{n=1}^N$};
			\node (ci2) [process, below=of ci1,align=center] %
			{Fit low-dimensional Gaussian \\ mixture distribution to $\bA\bxi$};
			\node (ci3) [process, below=of ci2] {Reformulated Chance Constraints};
			\node (ci4) [startstop, below=of ci3] %
			{Optimal Dispatch Decisions};

			\draw [arrow] (start) -- node[midway, above right, xshift=1pt] {} (ci1);
			\draw [arrow] (ci1) -- node[midway, right] {} (ci2);
			\draw [arrow] (ci2) -- (ci3);
			\draw [arrow] (ci3) -- (ci4);

			\node[draw=black, dashed, fit=(c1)(c2)(c3)(c4), inner sep=0.3cm, label={[align=center]above:\textbf{Classical Approach}\\Fit and then Transform}] {};
			\node[draw=black, dashed, fit=(ci1)(ci2)(ci3)(ci4), inner sep=0.3cm, label={[align=center]above:\textbf{Constraint-Informed Approach}\\Transform and then Fit}] {};
			
		\end{tikzpicture}
		\caption{\xreplaced{Comparison of the classical (fit and then transform) and constraint-informed (transform and then fit) approaches for reformulating chance constraints. The matrix $\bA$ denotes the low-dimensional linear projection identified from the constraint structure.}{Comparison of Classical and Constraint-Informed Approaches for Reformulating Chance Constraints}}
		\label{fig:combined_flowchart}
	\end{figure*}

    \xadded{We refer to this as the \emph{constraint-informed} approach because the projection $\bA$ is informed directly from the structure of the chance constraints. The matrix $\bA$ characterizes precisely how the raw uncertainty $\bxi$ enters and influences each constraint. %
	The approach applies to any linear chance-constrained model: one needs only to identify the low-dimensional projection $\bA\bxi$ and fit a GMM distribution to the transformations $\{\bA\bxi^{(n)}\}_{n=1}^N$ of the raw samples $\{\bxi^{(n)}\}_{n=1}^N$. %
	Because the distribution is fit to low-dimensional data, it is more accurate and efficient to learn compared to fitting $\bxi$ directly and then transforming its fitted parameters. We accordingly refer to the constraint-informed approach as \emph{transform and then fit}, in contrast to the classical \emph{fit and then transform}. Fig.~\ref{fig:combined_flowchart} illustrates this comparison.}
	
	For brevity, we present \xreplaced{the detailed reformulation steps for the upper-bound}{our ideas in the context of} constraints~\eqref{eq:cc_opf_power_max} and~\eqref{eq:cc_opf_line_flow_max} only. The \xreplaced{treatment of}{corresponding techniques for}~\eqref{eq:cc_opf_power_min} and~\eqref{eq:cc_opf_line_flow_min} \xreplaced{is}{are} entirely analogous. %
    
	\subsection{Classical Approach: Fit and then Transform}\label{sec:classical_approach}
	The vast majority of existing approaches, which we shall collectively refer to as the \emph{classical approach}, reformulate the chance constraints by first fitting a high-dimensional probability distribution to the raw forecast error data and then transforming the resulting distribution parameters to embed within the CC-OPF reformulation. To illustrate the main steps of the approach, consider $N$ observations of forecast errors:
	\begin{equation}\label{eq:data_samples}
		\{\bxi^{(n)}\}_{n=1}^N, \;\;\bxi^{(n)}\coloneqq \big(\xi_1^{(n)}, \xi_2^{(n)},\cdots, \xi_{|\mathcal{B}|}^{(n)} \big) \in\mathbb{R}^{|\mathcal{B}|},
	\end{equation}
	\begin{enumerate}
		\item Fit a multivariate Gaussian--or more generally, Gaussian Mixture--distribution, $\sum_{k=1}^{K} \hat{\pi}_k \mathcal{N}(\hat{\bmu}_k,\hat{\bSigma}_k)$, to the raw dataset $\{\bxi^{(n)}\}_{n=1}^N$. This is typically done using the Expectation-Maximization (EM) algorithm.
		
		\item Transform the fitted parameters to obtain the distributions of the random variables, $p_g(\bxi)$ and $f_l(\bxi)$, which appear in chance constraints~\eqref{eq:cc_opf_power_max}--\eqref{eq:cc_opf_line_flow_min}.
		Since these are affine functions of $\bxi$ (see \eqref{eq: actual_power}--\eqref{eq:net_power_inj}, \eqref{eq:flow}--\eqref{eq:flow_matrix}), we obtain:
		\begin{gather}\label{eq:gmm_distributed_pg_xi}
			p_g(\bxi) \sim\sum_{k=1}^{K} \hat{\pi}_k \mathcal{N}(\bar{p}_g - \alpha_g \bone^\top\hat{\bmu}_k,\; \alpha_g^2 \bone^\top\hat{\bSigma}_k\bone), \\
			f_l(\bxi) \sim\sum_{k=1}^{K} \hat{\pi}_k \mathcal{N}(
			E_{lk}(\balpha), V_{lk}(\balpha)), \\
			E_{lk}(\balpha) \coloneqq f_l^0+
			(\gamma_l(\balpha) \bone +
			\bh^{\text{wind}}_l)^\top \hat{\bmu}_{k}, \notag\\
			V_{lk}(\balpha) \coloneqq (\gamma_l(\balpha) \bone +
			\bh^{\text{wind}}_l)^\top \hat{\bSigma}_k (\gamma_l(\balpha) \bone +
			\bh^{\text{wind}}_l). \notag
		\end{gather}
		
		\item Reformulate chance constraints \eqref{eq:cc_opf_power_max} and \eqref{eq:cc_opf_line_flow_max} using the standard normal cumulative distribution function $\Phi$:
		\begin{gather}\label{eq:cc_classical_cdf_form}
			\sum_{k=1}^{K} \hat{\pi}_k \Phi\bigg(\frac{p_g^{\text{max}}-\bar{p}_g+\alpha_g\bone^\top\hat{\bmu}_k}{\alpha_g\sqrt{\bone^\top\hat{\bSigma}_k\bone}}\bigg) \geq 1-\epsilon, \\
			\sum_{k=1}^{K} \hat{\pi}_k \Phi\bigg(\frac{f_{l}^{\max}-E_{lk}(\balpha)}{\sqrt{V_{lk}(\balpha)}}\bigg) \geq 1-\epsilon.
		\end{gather}
		Reformulate the above as convex conic constraints using the method presented in Section~\ref{sec:pwl_reformulation}.
	\end{enumerate}

	\subsection{Constraint-Informed Approach: Transform and then Fit}
	The classical approach is an indirect way of targeting randomness in OPF constraints: it attempts to fit a high-dimensional joint distribution to $\{\bxi^{(n)}\}_{n=1}^N$ and hopes that the linear transformations of the estimators $\hat{\pi}_k, \hat{\bmu}_k,\hat{\bSigma}_k$ accurately represent the true mean and variance of stochastic power generation $p_g(\bxi)$ and line flow $f_l(\bxi)$. To address this issue, we propose our constraint-informed approach, which directly targets the randomness $\Omega,\boldeta_l$ that are present in constraints \eqref{eq:cc_opf_power_max}, \eqref{eq:cc_opf_line_flow_max}, respectively. %
	The main steps of the approach are:
	\begin{enumerate}
		\item Transform the data samples using \eqref{eq:1d_rand} and \eqref{eq: boldeta_l} to obtain data samples for the one- and two-dimensional system-wide and line~$l$ forecast errors, respectively:
		\begin{align}
			\{\Omega^{(n)}\}_{n=1}^N, &\;\;\Omega^{(n)}\coloneqq\sum_{i \in \mathcal B}  \xi_i^{(n)}\in\mathbb{R}, 
			\label{eq:Omega_construction} \\
			\{\boldeta_l^{(n)}\}_{n=1}^N, &\;\;\boldeta_l^{(n)}\coloneqq \left(\Omega^{(n)}, \sum_{i\in \mathcal{B}}H_{li}\xi_i^{(n)}\right) \in\mathbb{R}^2.
		\end{align}
		
		\item Fit low-dimensional GMMs to $\{\Omega^{(n)}\}_{n=1}^N$, $\{\boldeta_l^{(n)}\}_{n=1}^N$: 
		\begin{equation}\label{eq:gmm_distributed_1d}
			\Omega \sim\sum_{k=1}^{K} \hat{\beta}_k \mathcal{N}(\hat{m}_k, \hat{\sigma}_k^2),\;\boldeta_l\sim\sum_{k=1}^{K} \hat{\lambda}_k \mathcal{N}(\hat{\bnu_k},\hat{\bC_k}).
		\end{equation}
		\item Use the fitted parameters \xdeleted{to} to obtain the distributions of the random variables, $p_g(\bxi)$ and $f_l(\bxi)$, and to reformulate the chance constraints \eqref{eq:cc_opf_power_max} and \eqref{eq:cc_opf_line_flow_max}, respectively:
		\begin{gather}\label{eq:cc_cinform_cdf_form}
			\sum_{k=1}^{K} \hat{\beta}_k \Phi\bigg(\frac{p_g^{\text{max}}-\bar{p}_g+\alpha_g\hat{m}_k}{\alpha_g\hat{\sigma}_k}\bigg) \geq 1-\epsilon, \\
			\
			\sum_{k=1}^{K} \hat{\lambda}_k \Phi\Bigg(\frac{f_{l}^{\max}-f_l^0-
				(\gamma_l(\balpha), 1)^\top
				\hat{\bnu}_k}{\sqrt{
					(\gamma_l(\balpha), 1)^\top
					\hat{\bC}_k
					(\gamma_l(\balpha), 1)
			}}\Bigg) \geq 1-\epsilon.\label{eq:cc_cinform_cdf_form_line_flow}
		\end{gather}
		Reformulate the above as convex conic constraints using the method presented in Section~\ref{sec:pwl_reformulation}.
	\end{enumerate}
	
	When making dispatch decisions, system operators are primarily concerned with system-wide imbalances rather than individual unit deviations, thus directly modeling \(\Omega\) is also operationally meaningful. On the other hand, the construction of data samples for $\boldeta_l$ from raw forecast error data takes into consideration both aggregate system-wide forecast errors $\Omega$ and line-specific localized forecast errors $\Lambda_l$. Therefore, our constraint-informed approach efficiently captures both the global and local effects of forecast uncertainty, while avoiding high-dimensional statistical estimation.
	\xdeleted{%
	Fig.~\ref{fig:combined_flowchart} %
	shows the flowchart of the classical
	and constraint-informed approaches}. %
	
	\subsection{Benefits of Constraint-Informed Approach}\label{sec:benefits}
	The classical and constraint-informed approaches generally yield different chance constraint reformulations for a $K$-component Gaussian mixture model, due to differences in the estimated statistical parameters. 
	However, there is one special case where they coincide: namely, when we fit a Gaussian distribution to the forecast errors (where $K=1$), a model frequently used in existing work (e.g., see \cite{bienstock2014chance,roald2017chance,lubin2019chance}).
	
	In this case, it is well known that the classical chance constraint~\eqref{eq:cc_classical_cdf_form} can be written as a linear constraint:
	\begin{equation}
		\bar{p}_g - \alpha_g \left( \mathbf{1}^\top \hat{\bmu}_1 - \Phi^{-1}(1 - \epsilon) \sqrt{\mathbf{1}^\top \hat{\bSigma}_1 \mathbf{1}} \right)
		\leq p_g^{\max}.
		\label{eq: gaussian_power_reformulate_ftt_max}
	\end{equation}
	Similarly, the constraint-informed chance constraint \eqref{eq:cc_cinform_cdf_form} using a Gaussian model for $\Omega\sim \mathcal N(\hat{m}_1, \hat{\sigma}_1^2)$ simplifies to:
	\begin{equation}
		\bar{p}_g - \alpha_g (\hat{m}_1 - \Phi^{-1}(1 - \epsilon)  \hat{\sigma}_1 )
		\leq p_g^{\max}.
		\label{eq: gaussian_power_reformulate_ttf_max}
	\end{equation}
	It turns out that only in this special case of Gaussian distributions that are estimated using maximum likelihood estimation (MLE), we have $\bone^\top\hat{\bmu}_1 = \hat{m}_1$ and $\bone^\top\hat{\bSigma}_1\bone = \hat{\sigma}_1^2$ so that the reformulated linear constraints \eqref{eq: gaussian_power_reformulate_ftt_max} and \eqref{eq: gaussian_power_reformulate_ttf_max} are identical.
	This is because of the structure of the optimal maximum likelihood estimator of Gaussians (see Appendix~\ref{sec:appendix_mle}). An analogous argument holds for the line flow chance constraints as well.
	
\begin{figure}[htbp]
		\centering
		\includegraphics[width=0.85\linewidth]{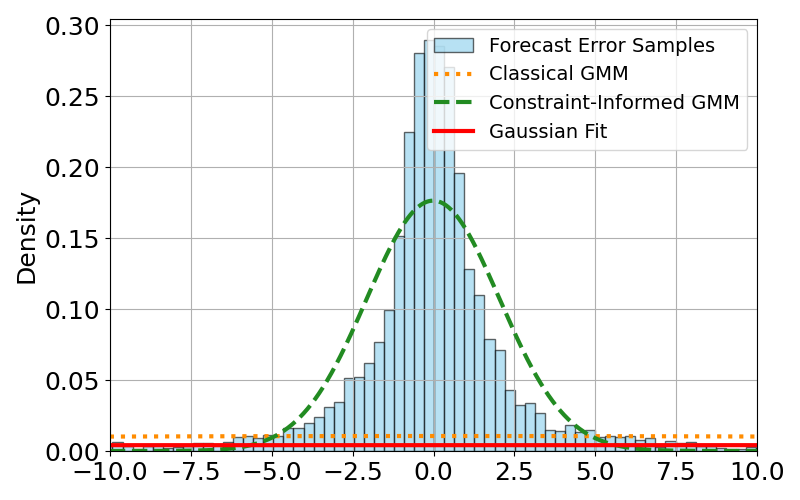}
		\caption{Aggregate forecast errors of five wind units using parameters in~\cite{hodge2011wind}}
		\label{fig:sec_3}
	\end{figure}

	In reality, however, the true distribution can deviate significantly from Gaussianity, especially when the forecast error data is multi-modal or has heavy Cauchy-like tails \cite{hodge2011wind,zhang2024efficient}.
	The example in Figure~\ref{fig:sec_3} and Table~\ref{tab:toy_statistics} shows that a Gaussian fit can yield a poor approximation even when aggregating forecast errors from as few as five wind units, each modeled as a Cauchy distribution with parameters from the literature~\cite{hodge2011wind}\footnote{The GMM is estimated using the \texttt{sklearn.mixture} Python package. %
		For each method, we perform ten runs of EM with different initializations and select the model with the lowest Bayesian Information Criterion (BIC).}.
	
	In the general GMM setting, the constraint-informed approach becomes particularly advantageous, as it retains a low-dimensional uncertainty even when the underlying distribution is no longer Gaussian.
	In this case, the classical and constraint-informed approaches yield different reformulated constraints due to differences in their statistical parameters.
	In fact, even if $\bxi, \Omega$ and $\boldeta_l$ are all estimated using GMMs with the same number of components $K$, we have
	$
	\hat{\pi}_k \neq \hat{\beta}_k
	$,
	$\bone^\top \hat{\bmu}_k \neq \hat{m}_k$, and
	$\bone^\top \hat{\bSigma}_k \bone \neq \hat{\sigma}_k^2
	$
	for any component $k$.
	This is illustrated in Table~\ref{tab:toy_statistics}, which compares the parameters of a 3-component GMM using both approaches on samples drawn from Figure~\ref{fig:sec_3}.
	In particular, the dominant component of the constraint-informed GMM exhibits significantly lower variance compared to that of the classical approach, also evidenced by the distinct density curves in Figure~\ref{fig:sec_3}.
	This can be attributed to the heuristic and non-global nature of the EM algorithm, which is typically used to estimate the model parameters.
	
	\begin{table}[htbp]
		\centering
		\caption{Fitted estimates of $\Omega$ on the dataset in Figure~\ref{fig:sec_3}}
		\label{tab:toy_statistics}
		\begin{tabular}{lrrr}
			\toprule
			\textbf{Classical} &  $\bone^\top\hat\bmu_k$ & $\bone^\top\hat\bSigma_k\bone$ & $\hat\pi_k$ \\ \midrule
			Component 1 & 0.24     & 1374.00     & 0.9994 \\
			Component 2 & 5366.81  & $\approx0.00$ & 0.0002 \\
			Component 3 & -1760.55 & 407031.26   & 0.0004 \\
			\midrule
			\textbf{Constraint-Informed} & $\hat m_k$ & $\hat\sigma_k^2$ & $\hat\beta_k$\\ \midrule
			Component 1 & -0.01    & 4.22        & 0.9079 \\
			Component 2 & 5366.81  & $\approx0.00$ & 0.0002 \\
			Component 3 & -4.89    & 30149.67    & 0.0919 \\
			\midrule
			\textbf{Gaussian MLE}           & 0.61     & 8536.01     & 1.0000 \\
			\bottomrule
		\end{tabular}
	\end{table}
	
	Another key drawback of the classical approach is that it relies on high-dimensional statistical fitting, where the number of parameters grows quadratically in the number of wind units due to covariance matrix estimation. In contrast, the constraint-informed approach fits $\Omega$ and $\boldeta_l$ directly to a one- and two-dimensional GMM, lowering the estimation burden.

	\subsection{Computational Considerations}
	In the constraint-informed approach, $\Omega$ serves as a common uncertainty term for all $|\mathcal{G}|$ constraints of the form \eqref{eq:cc_opf_power_min} and \eqref{eq:cc_opf_power_max}, but each of the $|\mathcal{L}|$ constraints of the form \eqref{eq:cc_opf_line_flow_min} and \eqref{eq:cc_opf_line_flow_max} requires a distinct random variable $\boldeta_l$, necessitating a higher number of model fittings compared to the classical approach. This introduces a tradeoff: the classical approach fits a single high-dimensional distribution, while the constraint-informed approach fits two-dimensional distributions, reducing the computational complexity per estimation task but increasing the number of required fits; see Table~\ref{tab:tradeoff}. 
	
	\begin{table}[htbp]
		\centering
		\caption{Classical versus Constraint-informed tradeoffs \label{tab:tradeoff}}
		\begin{tabular}{c|c|c}
			\toprule
			\textbf{Approach} & \textbf{Number of Model Fittings} & \textbf{Model Dimension}\\ \midrule
			Classical & $1$ & $|\mathcal{B}|$\\ \midrule
			Constraint-informed &   $1 + |\mathcal{L}|$\ & $\leq 2$ \\ \bottomrule
		\end{tabular}
	\end{table}
	
	The statistical fitting can be efficiently performed using standard software packages that provide implementations of the EM algorithm. %
	Notably, it can be done entirely off-line (e.g., day-ahead) once generator commitments are fixed.
	The online computational cost of solving the constraint-informed CC-OPF is identical to the classical approach, since the fitting does not interfere with real-time dispatch decisions.

	\subsection{GMM Chance Constraint Reformulation}\label{sec:pwl_reformulation}
	We use the constraint-informed estimates $\hat{\beta}_k$, $\hat{m}_k$, and $\hat{\sigma}_k^2$ for $\Omega$ to rewrite constraint~\eqref{eq:cc_opf_power_max} as \eqref{eq:cc_cinform_cdf_form}. 
	Unlike the Gaussian case,  there is no closed-form expression for the inverse cumulative distribution function (CDF) of a GMM. %
	As a result, the left-hand side expression in \eqref{eq:cc_cinform_cdf_form} %
	is both nonlinear and nonconvex, making direct reformulation difficult for optimization.
	
	To address this challenge, we adapt an existing method from the literature~\cite{fathabad2023asymptotically} that builds a piecewise linear (PWL) approximation of the standard normal CDF $\Phi$.
	The key idea is to discretize $\Phi$ over the nonnegative reals into $S$ linear segments, using breakpoints $t_0=0<t_1< \cdots<t_{S-1}$.
	\xreplaced{%
		Given a user-specified tolerance $\delta > 0$, these breakpoints are chosen optimally by \mbox{\cite[Algorithm~1]{fathabad2023asymptotically}} to minimize the number of segments~$S$ while ensuring that $\lVert\hat{\Phi}(x)-\Phi(x) \rVert\leq \delta$ for all $x\in [0,\infty)$. Thus, $S$ is not a free parameter but is determined by the choice of~$\delta$; a smaller $\delta$ yields a tighter approximation at the cost of more segments. We discuss how to set $\delta$ in Section~\ref{sec:numerical}.%
	}%
	{These breakpoints can be chosen optimally based on a user-specified tolerance $\delta$ \mbox{\cite[Algorithm~1]{fathabad2023asymptotically}} so that the linear approximation $\hat{\Phi}$ remains as close as possible to $\Phi$; namely, $\lVert\hat{\Phi}(x)-\Phi(x) \rVert\leq \delta, \forall x\in [0,\infty)$, with a minimal number of segments.%
	}
	For each segment defined over the interval $[t_{s-1}, t_s]$, we approximate $\Phi$ from \emph{below} with a linear function whose slope $a_s $ and intercept $b_s$ can be calculated as follows:
	\begin{align}
		a_s &= \frac{\Phi(t_s)-\Phi(t_{s-1})}{t_s-t_{s-1}}, &&\;\; \forall s=1,2,\ldots,S-1,
		\label{eq:pwl_slope} \\
		b_s &= \Phi(t_s)-a_st_s, &&\;\; \forall s=1,2,\ldots,S-1.
		\label{eq:pwl_intercept}
	\end{align}
	To handle the domain $[t_{S-1}, \infty)$ where $\Phi$ asymptotes, the slope and intercept of the rightmost horizontal segment are: 
	\begin{equation}
		a_S=0, \;\; b_S=\Phi(t_{S-1}). 
		\label{eq: pwl_endpoint}
	\end{equation}
	The resulting piecewise linear approximation $\hat{\Phi}$ is concave and can be expressed compactly as:
	\begin{equation}
		\hat{\Phi}(x)=\min_{s=1,\ldots, S}
		\left\{ a_sx+b_s \right\},
		\label{eq:pwl}
	\end{equation}
	The pointwise minimum ensures that the PWL approximation $\hat{\Phi}$ is a valid under-estimator for $\Phi$, so that the solution obtained by substituting $\Phi$ with $\hat{\Phi}$ in constraint~\eqref{eq:cc_cinform_cdf_form} will also be feasible for the original constraint~\eqref{eq:cc_opf_power_max}. \xadded{In other words, the PWL approximation always yields conservative (i.e., more reliable) solutions, regardless of the number of segments~$S$.}

	To ensure that discretizing over the nonnegative reals will suffice, we need to impose additional constraints \cite[Proposition~1]{fathabad2023asymptotically} that restrict the domain of $\Phi$ in~\eqref{eq:cc_cinform_cdf_form} to $[0,\infty)$. %
	\begin{equation}
		\bar{p}_g-\hat{m}_k\alpha_g\leq p_g^{\text{max}}, \;\;\;\forall k \in [K]. 
		\label{eq: mean_condition}
	\end{equation}
	Depending on the value of $\hat{m}_k$, this constraint can lead to an infeasible optimization model. This can be corrected by enforcing $\hat{m}_k = 0$ during fitting using a modified EM algorithm \cite{yi2024discrete}, effectively making constraint~\eqref{eq: mean_condition} non-binding.
	
	Substituting $\Phi$ with its PWL approximation $\hat{\Phi}$ in~\eqref{eq:cc_cinform_cdf_form} yields:
	\begin{align*}
		\sum_{k=1}^{K} \hat{\beta}_k %
		\min_{s=1,\ldots, S}
		\left\{
		a_s\bigg(\frac{p_g^{\text{max}}-\bar{p}_g+\hat{m}_k\alpha_g}{\hat{\sigma}_k\alpha_g}\bigg)+b_s
		\right\}
		\geq 1-\epsilon.
	\end{align*}
	We can multiply both sides by (the nonnegative term) $\alpha_g$ to eliminate nonlinearity in the constraint.
	Similarly, we can introduce \xreplaced{a continuous}{an} auxiliary variable $M_{gk}^1$ for each \xreplaced{component $k \in [K]$}{piecewise linear segment} and each $g \in \mathcal G$ to eliminate the minimum operator\xadded{; we emphasize that no binary variables are needed}.
	\begin{gather}
		\sum_{k=1}^{K} \hat{\beta}_k M_{gk}^1 \geq (1-\epsilon)\alpha_g,
		\label{eqref:M1k} \\
		M_{gk}^1\leq a_s\bigg(\frac{p_g^{\text{max}}-\bar{p}_g+\hat{m}_k\alpha_g}{\hat{\sigma}_k}\bigg)+b_s\alpha_g, \;\forall s \in [S].
		\label{eqref:M1k_aux}
	\end{gather}
	
	The reformulation of the line flow constraint~\eqref{eq:cc_cinform_cdf_form_line_flow} follows a similar idea to that of the power generation constraint~\eqref{eq:cc_cinform_cdf_form}, where we need a constraint analogous to \eqref{eq: mean_condition}:
	\begin{equation}
		f_l^0 + (\gamma_l(\balpha), 1)^\top\hat{\bnu}_k\leq f_l^{\max},\;\forall k \in [K].
		\label{eq: mean_condition_flow}
	\end{equation}
	Convexity of the reformulation, however, requires the additional assumption that the component-specific covariance matrices of $\boldeta_l$ share the same covariance matrix\footnote{This is supported by standard statistical software; e.g., \texttt{spherical} or \texttt{tied} covariance in the \texttt{sklearn.mixture} Python package. \xadded{In our experiments, the \texttt{tied} structure achieved nearly identical log-likelihoods to unrestricted covariances while using fewer parameters, and this advantage is particularly pronounced when data are scarce.}}: $\hat{\bC}_k = \tau_k^2 \hat{\bC}_0$. \xadded{%
	Relaxing this assumption for $\boldeta_l$ would require introducing binary variables to handle arbitrary component covariances~\cite{dey2025solving}, destroying convexity and real-time applicability of the method.}
	Utilizing techniques similar to \eqref{eqref:M1k} and \eqref{eqref:M1k_aux}, we obtain:
	\begin{gather}
		\sum_{k=1}^{K} \hat{\lambda}_k M_{lk}^3 \geq (1-\epsilon) \delta_l,
		\label{eqref:M3k} \\
		M_{lk}^3\leq a_s\left( \frac{f_l^{\text{max}}-f_l^0-
			(\gamma_l(\balpha), 1)^\top
			\hat{\bnu}_k
		}{\tau_k} \right) + b_s \delta_l,  \;\forall s \in [S].
		\label{eqref:M3k_aux}
	\end{gather}
	Here, $\delta_l \geq 0$ is an auxiliary decision variable satisfying the convex second-order cone constraint:
	\begin{equation}
		\delta_l^2 \geq (\gamma_l(\balpha), 1)^\top \hat{\bC}_0 (\gamma_l(\balpha), 1).
		\label{eq: delta_l}
	\end{equation}
	A detailed derivation can be found in~\cite[Proposition~1]{fathabad2023asymptotically}.
	The resulting CC-OPF problem, with reformulated power generation and line flow limits, is a second-order cone convex optimization problem that can be solved using commercial solvers. A complete formulation can be found in Appendix~\ref{sec:appendix_reformulation}.
	
	\section{Computational Experiments}\label{sec:numerical}
	
	We use Julia~1.5.3 with PowerModels.jl \cite{coffrin2018powermodels} and the Gurobi solver. %
	All runs were performed on
	a personal MacBook Air (M2 chip with 8-core CPU %
	and 16GB RAM). The code and data files for reproducing the results are at \href{https://github.com/Subramanyam-Lab/Constraint_Informed_CCOPF/}{https://github.com/Subramanyam-Lab/Constraint\_Informed\_CCOPF/}.%
	
	\subsection{Test Systems and Datasets}
	We perform experiments on a modified IEEE~118 bus test case. In particular, we replace 10 out of 19 conventional generators %
	with wind units that don't incur any generation cost. %
	\xadded{To assess scalability, we also consider a larger 2746-bus Polish system in Section~\ref{sec:scalability}.}
	We set the chance constraint risk threshold to $\epsilon = 0.05$. The accuracy of the piecewise linear approximation of $\Phi$ is set to \xreplaced{$\delta = 0.1\epsilon$}{$0.002$, which results in $S = 10$ segments being used}.%
		\footnote{\xadded{This choice can be explained as follows. Since %
		$0\leq \Phi(x)-\hat\Phi(x)\leq\delta$ for all $x\geq 0$, enforcing the chance constraints with $\hat\Phi$ in place of $\Phi$ can be shown to introduce at most $\delta$ additional conservatism compared to the original constraint (since the mixture weights sum to one). The factor of $0.1$ ensures that this conservatism is at least one order of magnitude smaller than~$\epsilon$.}} \xadded{With $\epsilon=0.05$, this gives $\delta = 0.005$, resulting in $S = 6$ PWL segments}\footnote{\xadded{Reducing $\delta$ results in more PWL segments $S$ that increase computational time but also provide a tighter approximation of the Gaussian CDF and hence a decrease in cost and solution conservatism. We find that tightening the tolerance to $\delta = 0.002$ ($S = 10$) and $\delta = 0.001$ ($S = 13$) increased solution time without any meaningful change in out-of-sample risk.}}\xadded{.}
	
	We experiment with three datasets $\{\bxi^{(n)}\}_{n=1}^N$ of forecast errors: synthetic Gaussian-distributed errors (Synthetic-G), synthetic Cauchy-distributed errors (Synthetic-C), and real-world NordPool data \cite{websiteNordPool}. 
	\xadded{The choice of the Cauchy distribution is illustrative; the proposed framework applies to any continuous distribution since GMMs are universal approximators \cite{mclachlan2019finite}. The forecast errors themselves need not follow a GMM and the latter is only used as a statistical approximation of the empirical data.}
	For the synthetic cases, we generate ten independent datasets of forecast errors using distributional parameters\footnote{%
		Gaussian: $\mu=-0.024, \sigma=0.036$ and Cauchy: $x_0=0, \gamma=0.02$.} adapted from existing literature \cite{song2024chance,hodge2011wind}.
	Each dataset consists of 10,000 samples, with $N = 8,000$ samples used for statistical estimation and subsequent optimization, while the remaining 2,000 for evaluating out-of-sample risk.
	
	For the NordPool case, we use real 15-minute wind power production and intraday forecast data from seven locations in the CWE region: 50Hz, AMP, AT, FR, PL, TBW, and TTG. Among these, 50Hz, FR, and TTG include both onshore and offshore wind data, while the others have only onshore data. To ensure consistency with the synthetic cases, we treat onshore and offshore wind generation as separate units, resulting in a total of ten wind units.
	For each wind unit, we normalize forecast errors: $(\text{actual} - \text{forecast})/\text{actual}$.
	We collect 15 consecutive days (February 26 - March 12, 2025) of NordPool data, yielding 1,440 total samples, with 80\% used for statistical estimation and optimization and the remaining 20\% for out-of-sample risk evaluation.
	
	\subsection{Goodness-of-Fit}\label{sec:goodness_of_fit}
\begin{figure*}[htbp]
		{\label{Fig1:a} \includegraphics[width=0.32\linewidth]{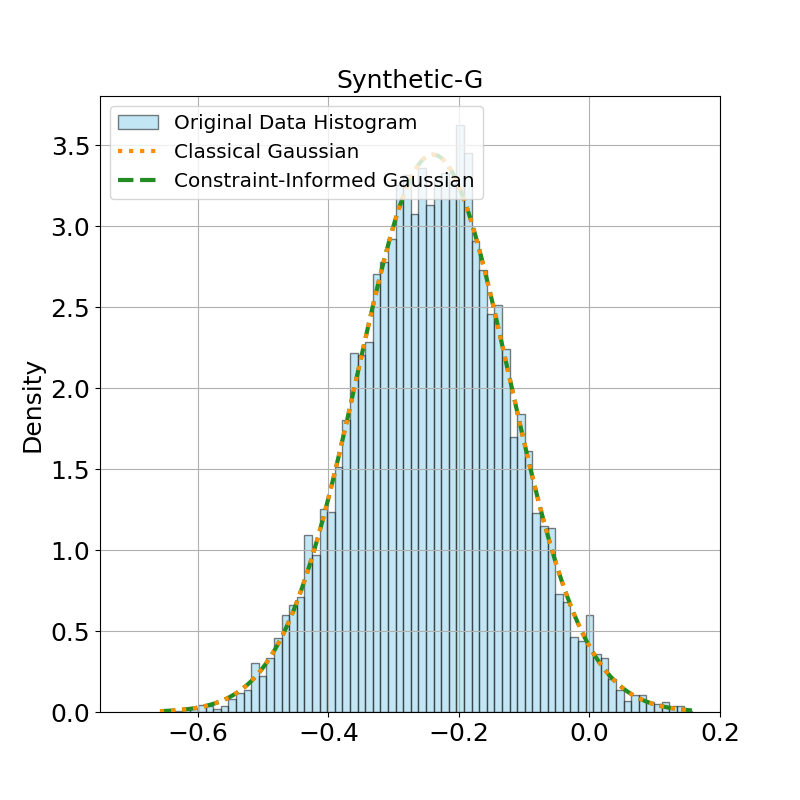}}%
		\hfill
		{\label{Fig1:b} \includegraphics[width=0.32\linewidth]{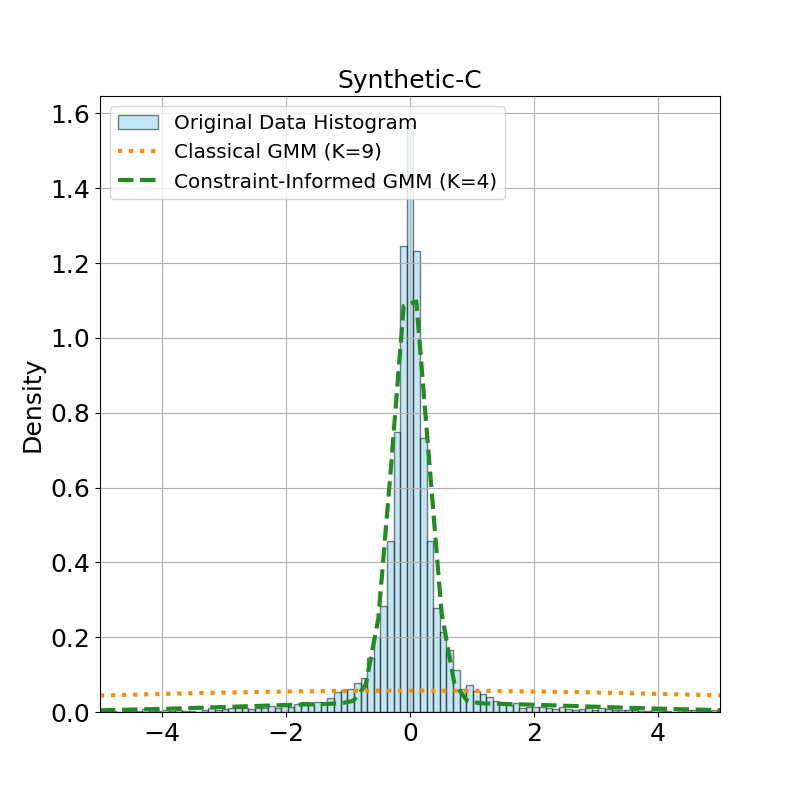}}
		\hfill
		{\label{Fig1:c} \includegraphics[width=0.32\linewidth]{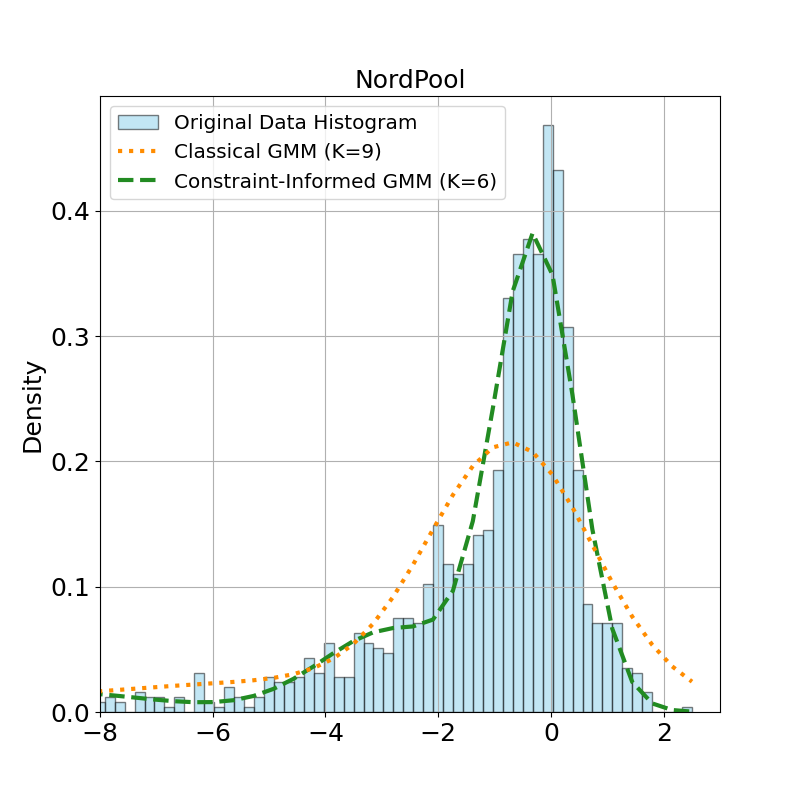}}%
		\caption{Aggregate errors and best-fit probability density curves estimated using the classical and constraint-informed approaches.} \label{Fig1:fgdd}
	\end{figure*}

	We assess the statistical goodness-of-fit of the classical and constraint-informed approaches through both visual and numerical comparisons. %
	For visual comparisons, we plot the following three quantities in Figure~\ref{Fig1:fgdd}.
	\begin{enumerate}
		\item True samples for aggregate system-wide forecast errors, $\{\Omega^{(n)}\}_{n=1}^N$, obtained using \eqref{eq:Omega_construction}.
		\item Density of $\Omega$ obtained using the classical `fit and then transform' approach, see Section~\ref{sec:classical_approach} for details.
		\item  Density
		of $\Omega$ using constraint-informed approach~\eqref{eq:gmm_distributed_1d}.
	\end{enumerate}
	
	For numerical comparisons, we compute the \xreplaced{Bayesian Information Criterion (BIC) scores of the GMMs estimated using both approaches for $K=1,2,\dots, 10$ Gaussian components. For each $K$, we estimate the GMM using 10 different EM initializations and record the lowest BIC. The BIC balances goodness-of-fit with model complexity: a lower BIC indicates a better fit relative to model size \cite{yi2024discrete}. Table~\ref{tab:bic_Cau} reports the lowest BIC and associated optimal $K$ for each dataset and approach. The corresponding density curves are plotted in Figure~\ref{Fig1:fgdd}.}{log-likelihood of the two densities to quantify the similarity of the estimated distributions to the empirical aggregate forecast errors.}%
		\begin{table}[htbp]
			\centering
			   \caption{\xreplaced{Best BIC scores and optimal number of components $K$ for the aggregate forecast error $\Omega$ (lower BIC is better)}{Log-likelihood of $\Omega$ (higher is better)}}%
			\label{tab:bic_Cau}
			\begin{tabular}{c|c|c}
				\toprule
				\textbf{Dataset}
				& \textbf{Classical} & \textbf{Constraint-Informed}\\ \midrule
				Synthetic-G& $-14761.09\ (K=1)$ & $-14716.09\ (K=1)$ \\ \midrule
				Synthetic-C & $44009.11\ (K=9)$ & $8986.49\ (K=4)$ \\ \midrule
				NordPool &  $5721.74\ (K=9)$  & $5517.40\ (K=6)$ \\ 
				\bottomrule
			\end{tabular}
		\end{table}
	\xdeleted{
		A higher log-likelihood indicates a better fit of the distribution to the observed data. %
		Table~\ref{tab:bic_Cau} reports the best log-likelihood values, obtained from ten different initializations of the EM algorithm across ten different splits of each dataset. %
		The corresponding density curves are plotted in Figure~\ref{Fig1:fgdd}.}

In the Synthetic-G dataset, where forecast errors are normally distributed, we find that the classical and constraint-informed approaches \xreplaced{select $K=1$ with nearly identical BIC}{yield equal log-likelihood} values and \xdeleted{identical} overlapping density curves, %
validating the MLE \xadded{equivalence} argument from Section~\ref{sec:benefits}.
\xdeleted{Unsurprisingly, the positive log-likelihood value indicates a good Gaussian fit in this case.}

	The results for the Synthetic-C dataset highlight the limitations of the classical approach. We believe this can be explained as follows: since the Cauchy distribution has undefined first and second moments, the conditions for the Central Limit Theorem do not apply. As a result, summing Cauchy-distributed errors does not approximate a normal distribution. The classical method, which fits a high-dimensional distribution and then reduces it to one dimension, %
	overemphasizes the heavy tails by estimating a high-variance Gaussian component. \xdeleted{It is possible to improve the log-likelihood %
	using a modified EM algorithm~\mbox{\cite{yi2024discrete}}; we present this detail in Appendix~\ref{sec:appendix_fixedEM}.}   
	
	The failure of the classical approach is evidenced in the middle and right-hand side plots of Figure~\ref{Fig1:fgdd}\xreplaced{: even with more Gaussian components, the classical approach achieves substantially higher BIC scores than the constraint-informed approach, and its}{, where} density curves deviate significantly from the true histograms of the Synthetic-C and NordPool datasets. In contrast, the constraint-informed approach, which directly fits a distribution to one-dimensional aggregated data, provides a much better match to the true samples. This is also supported by the \xreplaced{BIC scores}{log-likelihood values} in Table~\ref{tab:bic_Cau}. %
	\xadded{This improvement in fitting is also achieved with a significantly smaller number of mixture components. For the Synthetic-C dataset, the constraint-informed approach requires only 4 Gaussian components, whereas the classical approach requires 9. Previous work~\cite{wang2017chance} reported 7--9 Gaussian components for a GMM fit of $\bxi$ in a 39-bus system; our results both validate this observation and demonstrate that the proposed approach can achieve better accuracy with fewer components.}%

	\subsection{Out-of-Sample Risk and System Reliability}
	\xadded{%
	As noted in Section~\ref{sec:pwl_reformulation}, unconstrained EM may produce component mean estimates that violate~\eqref{eq: mean_condition} and~\eqref{eq: mean_condition_flow}, rendering the optimization model infeasible. To prevent this, all optimization-related results in this and subsequent sections use the FixedEM algorithm~\cite{yi2024discrete}, which constrains every component mean to zero for both the classical ($\bxi$) and the constraint-informed approach ($\Omega,\boldeta_l$). For each $\boldeta_l$ fitting, we fix $K=3$ and fit under both tied and spherical covariance structures with 10 initializations each; because the zero-mean constraint reduces the effective model complexity, BIC does not favor additional components beyond $K=3$.}

\begin{figure}[htbp]
		\centering
		\includegraphics[width=0.85\linewidth]{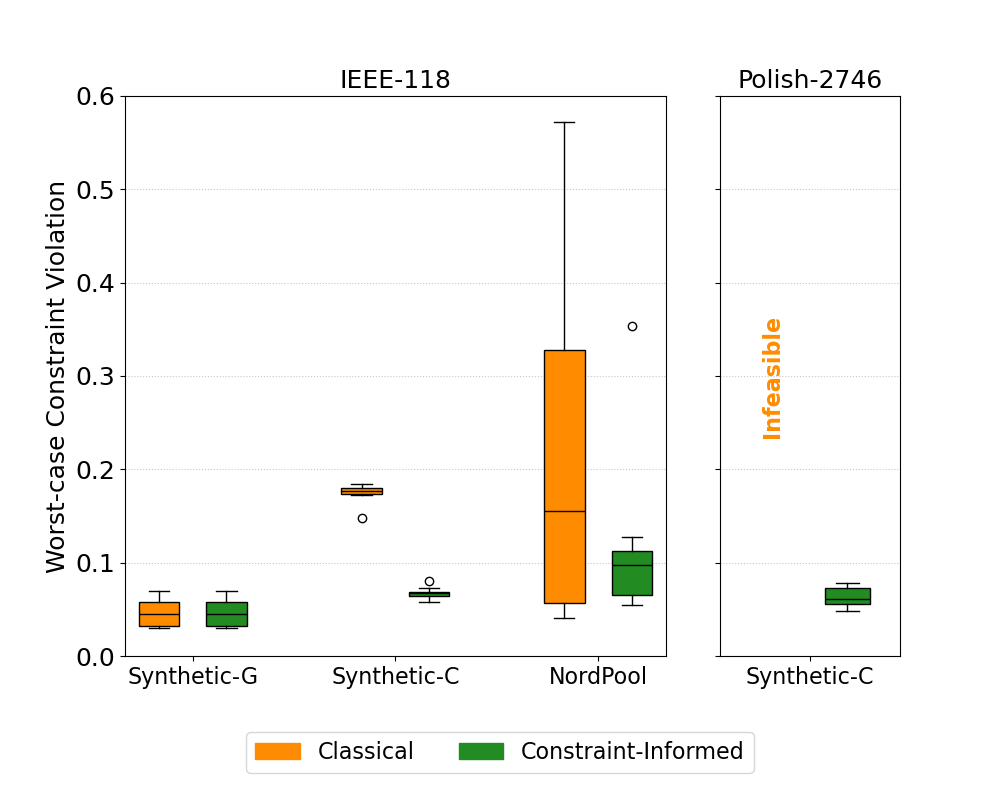}
		\caption{Out-of-sample worst-case constraint violations \xadded{across ten runs. The right panel corresponds to the test system discussed in Section~\ref{sec:scalability}, where the classical approach produced infeasible optimization models in all runs}.}
		\label{fig:sec_4c}
	\end{figure}

	To evaluate the quality of \xadded{CC-OPF} solutions obtained, we use 20\% of the dataset as a holdout, and denote it as $\mathcal{D}$. 
	The empirical violation of a single chance constraint (i.e., generation or line flow limit) is then defined as \cite{roald2017chance}:
	\begin{equation*}
		\rho_j = \frac{|\{\bxi^{(i)} \in \mathcal{D} \;|\; \text{constraint } j\;\text{is violated under } \bxi^{(i)}\}|}{|\mathcal{D}|}.
		\label{eq: rho_j}
	\end{equation*}
	If $\rho_j > \epsilon$, then the corresponding constraint does not meet the desired violation level, whereas $\rho_j \leq \epsilon$ indicates that the system remains secure as far as that particular generator or transmission line is concerned. Figure~\ref{fig:sec_4c} shows the worst-case violation, $\max_{j\in \mathcal{C}} \rho_j$, averaged across ten runs on each dataset, where $\mathcal{C}$ is the set of all constraints in \eqref{eq:cc_opf_power_min}-\eqref{eq:cc_opf_line_flow_max}.

	In the Synthetic-G dataset, the \xreplaced{negative BIC score}{positive log-likelihood} in %
	Table~\ref{tab:bic_Cau}
	indicates that the obtained Gaussian parameters are a good fit to the forecast errors. This directly translates to the constraint violations being below the targeted $\epsilon=0.05$. 
	
	In the Synthetic-C dataset, we observe that both approaches yield worst-case constraint violations that exceed $\epsilon$. The violations in the classical approach average roughly \xreplaced{$0.2$}{$0.5$}, whereas the constraint-informed approach achieves much smaller violations, averaging \xreplaced{slightly below $0.1$}{around $0.1$} with lower overall variance across the ten runs.
    
	Similar patterns can be seen in the NordPool dataset, where constraint-informed estimation consistently results in lower violations and variability compared to the classical approach. \xadded{The NordPool dataset exhibits both skewness and a limited number of samples, leading to greater variability across runs even under the constraint-informed approach. We also note that constraint-informed violations are higher on NordPool than on Synthetic-C, suggesting that enforcing zero-mean components may be less effective when the underlying distribution is significantly skewed.}

	\xadded{Overall, these results demonstrate that the constraint-informed dimensionality reduction consistently improves out-of-sample reliability: by fitting GMMs directly to the low-dimensional variables that enter each chance constraint, the EM algorithm better captures heavy-tailed (Synthetic-C) and skewed (NordPool) forecast errors, translating the improved goodness-of-fit into lower and less variable constraint violations.} \xdeleted{In fact, even when the classical approach enforces zero means, it still requires estimating a high-dimensional covariance matrix. This led to 3 infeasible models for the Synthetic-C dataset and 2 infeasible models for the NordPool dataset. In contrast, the proposed constraint-informed low-dimensional fitting produced no infeasible models in either case. }

	\xdeleted{We highlight that both approaches can sometimes produce statistical parameters that lead to infeasible optimization models, as shown in Figure~\ref{fig:sec_4c2}.
	The constraint-informed estimation, however, results in fewer infeasible models.
	We find that the infeasibility occurs because the estimated parameters violate constraints~\eqref{eq: mean_condition} and~\eqref{eq: mean_condition_flow}. %
	This infeasibility can be resolved using a modified FixedEM algorithm \mbox{\cite{yi2024discrete}} to constrain the means of the GMM components at zero (see Appendix~\ref{sec:appendix_fixedEM}).}

	\subsection{Computational Time}\label{sec:comp_time}
	\xreplaced{Table~\ref{tab:1} reports GMM fitting times on the NordPool dataset using the FixedEM algorithm. For $\bxi$ and $\Omega$, we sweep $K \in \{1,\dots,10\}$ with 10 random initializations per $K$ and select the best model by BIC. For each $\boldeta_l$, we fix $K=3$ and fit under both tied and spherical covariance structures with 10 initializations each. The per-line fitting time is consequently much smaller than that of $\Omega$, since fewer candidate models are evaluated.}{We compare total computation times for GMM estimation ($K=3$) on the NordPool dataset. Table~\ref{tab:1} reports the total time across 10 runs of EM, each with a different random initialization (and under both spherical and tied covariances in case of $\boldeta_l$), and then selecting the best model using the Bayesian Information Criterian (BIC) score.}
	\xreplaced{Due to the small size of the network, the optimization time for both approaches is relatively small. For the constraint-informed approach, the total computational time is 19.0 seconds, which is suitable for practical real-time deployment.}{We find that the computational times are small and suitable for practical real-time deployment.}
	\xadded{Moreover, the per-line fits for $\boldeta_l$ are independent and can be trivially parallelized.}
	\xdeleted{Unsurprisingly, the optimization time is identical for both approaches.}
	\begin{table}[htbp]
		\centering
		\caption{Computational time comparisons on the IEEE 118-Bus System}
		\label{tab:1}
		\begin{tabular}{c|c|c}
			\toprule
			& \textbf{Classical} & \textbf{Constraint-Informed}\\ \midrule
			Statistical Fitting & \xreplaced{2.81s}{0.04s} for $\bxi$ & \xreplaced{2.61s}{0.05s} for $\Omega$ and 0.09s per $\boldeta_l$\\ \midrule
			Optimization &  \xreplaced{0.34s}{0.03s}  & \xreplaced{0.43s}{0.03s} \\ \bottomrule
		\end{tabular}
	\end{table}

    \subsection{Scalability}\label{sec:scalability}
    \xadded{To evaluate the scalability of the constraint-informed approach, we consider the 2746-bus Polish 2003--2004 winter peak case~\cite{zimmerman2010matpower}. Following~\cite{bienstock2014chance}, we simulate 20\% wind penetration (of total load), evenly distributed across 50 wind units co-located with the 50 largest controllable generators. We use the Synthetic-C dataset where each wind unit is assigned Cauchy-distributed forecast errors.}\footnote{\xadded{Cauchy parameters: $x_0=0,\gamma=0.01$. The scale parameter~$\gamma$ is reduced from $0.02$ in the IEEE-118 case to moderate the forecast error magnitude for this larger network.}}

\begin{figure}[htbp]
		\centering
		\includegraphics[width=0.85\linewidth]{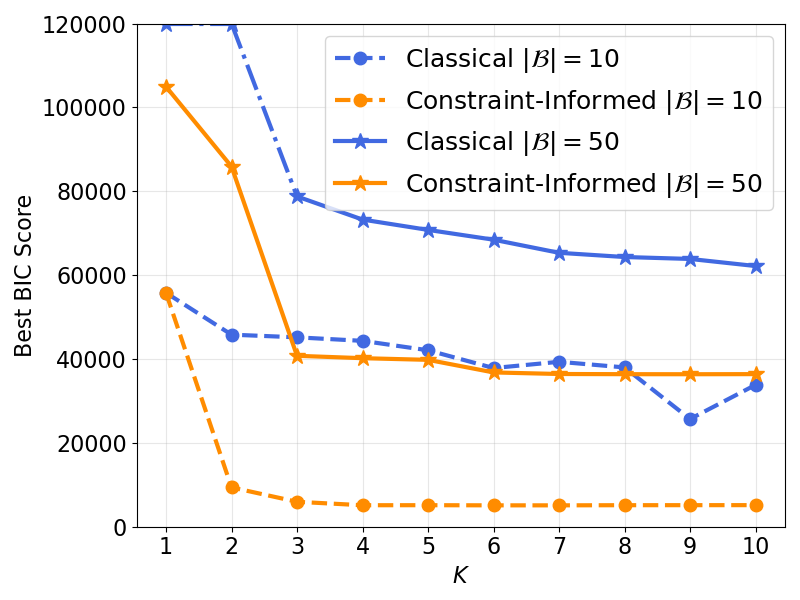}
		\caption{\xadded{Best BIC score versus $K$ for the aggregate forecast error~$\Omega$ at different uncertainty dimensions on the Polish 2746-bus system.}}
		\label{fig:bic_scalability}
	\end{figure}

	\xadded{Figure~\ref{fig:bic_scalability} plots the best BIC score as a function of the number of Gaussian components~$K$ for both approaches as the uncertainty dimension increases from 10 to 50. As in Section~\ref{sec:goodness_of_fit}, all goodness-of-fit comparisons in Figure~\ref{fig:bic_scalability} use classical EM to isolate the effect of dimensionality on statistical accuracy. The constraint-informed approach consistently achieves lower (better) BIC values across all~$K$, and its 50-dimensional curve nearly overlaps with the classical fit at 10 dimensions. Moreover, the constraint-informed BIC stabilizes around $K=3$ or $4$ even at high dimensions, indicating that a small number of components suffices to capture the distribution of~$\Omega$. In contrast, the classical approach does not exhibit a clear BIC minimum within the examined range; at $K\leq2$ in 50 dimensions, the fitted GMM assigns negligible density to tail observations, causing a collapse in the log-likelihood and a sharp increase in BIC.}

	\xadded{Table~\ref{tab:Polish_time} reports fitting and optimization times on the Polish system using the FixedEM algorithm, following the same procedure as in Section~\ref{sec:comp_time}. The optimization times are averaged across all 10 runs irrespective of infeasibility status. As the uncertainty dimension grows from 10 to 50, the classical fitting time increases substantially, whereas the constraint-informed fitting time for~$\Omega$ remains nearly unchanged since it always performs univariate estimation. The main computational cost of the constraint-informed approach lies in the per-line fits for~$\boldeta_l$; however, the per-line fitting time is roughly constant regardless of network size, and these fits are independent and can be fully parallelized. Even when performed sequentially, the total fitting and optimization time is 531.0 seconds, which is still within a typical OPF dispatch window. On the optimization side, the classical approach yields infeasible models due to poor high-dimensional parameter estimates, while the constraint-informed approach produces a feasible solution within minutes. These results confirm that the constraint-informed approach scales to large systems, maintaining both statistical accuracy and computational tractability.}

    \begin{table}[htbp]
		\centering
		\caption{\xadded{Computational time comparisons on the Polish 2746-bus system}}
		\label{tab:Polish_time}
		\begin{tabular}{c|c|c}
			\toprule
			& \textbf{Classical} & \textbf{Constraint-Informed}\\ \midrule
			Statistical Fitting & 12.27s for $\bxi$ & 3.01s for $\Omega$ and 0.13s per $\boldeta_l$\\ \midrule
			Optimization &  183.97s  & 101.68s \\ \bottomrule
		\end{tabular}
	\end{table}

    \xadded{The out-of-sample worst-case constraint violation for the constraint-informed approach is also illustrated in Figure~\ref{fig:sec_4c}. The stability of the low-dimensional fitting indicates that the worst-case violation remains largely unchanged as the network scales from 118 buses to 2746 buses. In contrast, the classical approach yields infeasible models across all ten runs, due to estimation challenges in fitting 50-dimensional GMMs. }
	\section{Conclusions}
	We study the interplay between statistics and optimization in the chance-constrained optimal power flow (CC-OPF) problem under uncertainties arising from wind power forecast errors. Our proposed \xreplaced{constraint-informed}{constrained-informed} approach uses the OPF constraints to isolate the true relevant uncertainties to two sources of randomness: a one-dimensional aggregate system-wide forecast error and a two-dimensional line-specific flow error. By reducing dimensionality, our approach integrates statistical estimation with chance-constrained optimization, thus providing system operators with information to make more reliable dispatch decisions. 
	
	When paired with GMM models, the constraint-informed approach can help alleviate the effects of heavy-tailed and skewed forecast errors that are present in both synthetic and real-world NordPool datasets. In particular, constraint-informed GMM increases the estimation accuracy of aggregate system-wide forecast errors, while also allowing different lines to have specific covariance structures. The increase in estimation accuracy translates to improved out-of-sample risk, without sacrificing total computational time.

	\xadded{A limitation of the current approach is that it fits separate low-dimensional distributions for each constraint-relevant projection rather than constructing a single globally coherent low-dimensional model. While this does not affect the enforcement of individual (or even two-sided) chance constraints, it would need to be addressed in settings that require joint chance constraints coupling multiple generators or lines. Extending the constraint-informed framework to such joint formulations, as well as to time-coupled problems such as unit commitment, is a promising direction for future work.}

	\appendices
	\section{Equivalence of Classical and Constraint-Informed Gaussian MLE}\label{sec:appendix_mle}
	Consider $N$ data samples $\bx^{(j)}\in\mathbb R^D$, for $j=1,\dots,N$.  
	
	In the classical approach, maximum likelihood estimation (MLE) for the Gaussian parameters results in the estimates:
	\begin{equation}\label{eq:mle_classical}
		\hat{\bmu}
		=\frac{1}{N}\sum_{j=1}^N\bx^{(j)}, 
		\;\;
		\hat\bSigma
		=\frac{1}{N}\sum_{j=1}^N\bigl(\bx^{(j)}-\hat{\bmu}\bigr)
		\bigl(\bx^{(j)}-\hat{\bmu}\bigr)^\top.
	\end{equation}
	
	In the constraint-informed approach, we instead transform the data first to obtain the lower-dimensional samples:
	\begin{equation}\label{eq:Omega_def}
		\Omega^{(j)}=\bone^\top\bx^{(j)},\quad j=1,\dots,N.
	\end{equation}
	We then use MLE to find estimates for the Gaussian parameters for
	$\Omega$ to obtain:
	\begin{equation}\label{eq:mle_constraint-informed}
		\hat m
		=\frac{1}{N}\sum_{j=1}^N\Omega^{(j)}, 
		\;\;
		\hat\sigma^2
		=\frac{1}{N}\sum_{j=1}^N\bigl(\Omega^{(j)}-\hat m\bigr)^2.
	\end{equation}
	
	Alternatively, we could have transformed the classical MLE parameters using the property of linear transformation of Gaussians: if $\bx \sim \mathcal{N}(\hat{\bmu}, \hat{\bSigma})$, then $\bA \bx \sim \mathcal{N}(\bA \hat{\bmu},\,
	\bA\hat\bSigma\,\bA^\top )$ for any matrix $\bA$.
	Let $\bA=\bone^\top$ and use \eqref{eq:mle_classical}--\eqref{eq:mle_constraint-informed} to get:
	\[
	\bone^\top\hat{\bmu}
	=\frac1N\sum_{j=1}^N\bone^\top\bx^{(j)}
	=\frac1N\sum_{j=1}^N\Omega^{(j)}
	=\hat m,
	\]
	\[
	\bone^\top\hat\bSigma\,\bone
	=\frac1N\sum_{j=1}^N\bigl(\bone^\top(\bx^{(j)}-\hat{\bmu})\bigr)^2
	=\frac1N\sum_{j=1}^N(\Omega^{(j)}-\hat m)^2
	=\hat\sigma^2.
	\]
	We thus find that both approaches yield identical parameter estimates whenever a Gaussian model is used for fitting data.

	\section{Constraint-Informed CC-OPF Reformulation}\label{sec:appendix_reformulation}
	\begin{small}
		\begin{align*}
			& \mathop{\text{minimize}}_{\bar{\bp}, \balpha,\bm{\delta}, M^1, M^2, M^3, M^4} \;  \eqref{eq: obj_final} \\
			\text{s.t.} \quad 
			&\eqref{eq:cc_opf_affine}, \;\eqref{eq:cc_opf_power_balance},\\
			& \eqref{eq: mean_condition}, \eqref{eqref:M1k},\eqref{eqref:M1k_aux}, \;\forall g\in\mathcal{G},\\
			&\bar{p}_g-\hat{m}_k\alpha_g\geq p_g^{\text{min}}, \;\forall k \in [K],\;\forall g\in\mathcal{G}, \\
			&\sum_{k=1}^{K} \hat{\beta}_k M_{gk}^2 \geq (1-\epsilon)\alpha_g, \;\forall g\in\mathcal{G}, \\
			&M_{gk}^2\leq a_s\bigg(\frac{\bar{p}_g-\hat{m}_k\alpha_g-p_g^{\text{min}}}{\hat{\sigma}_k}\bigg)+b_s\alpha_g, \;\forall s \in [S], \;\forall g\in\mathcal{G}, \\
			& \eqref{eq: mean_condition_flow},\eqref{eqref:M3k}, \eqref{eqref:M3k_aux}, \eqref{eq: delta_l}, \;\forall l\in\mathcal{L}, \\
			& f_l^0 + (\gamma_l(\balpha), 1)^\top\hat{\bnu}_k\geq -f_l^{\max},\;\forall k \in [K],\;\forall l\in\mathcal{L},\\
			& \sum_{k=1}^{K} \hat{\lambda}_k M_{lk}^4 \geq (1-\epsilon) \delta_l, \;\forall l\in\mathcal{L}, \\
			&M_{lk}^4\leq a_s\left( \frac{f_l^{\text{max}}+f_l^0+
				(\gamma_l(\balpha), 1)^\top
				\hat{\bnu}_k}{\tau_k} \right) + b_s \delta_l,  \\
			&\hspace{0.6\linewidth}\;\forall s \in [S], \;\forall l\in\mathcal{L}.
		\end{align*}
	\end{small}

	\bibliographystyle{IEEEtran}
	\bibliography{references}
	
	\begin{center}
		\scriptsize \framebox{
			\parbox{3.2in}{
				Government License (will be removed at publication):
				The submitted manuscript has been created by UChicago Argonne, LLC,
				Operator of Argonne National Laboratory (``Argonne").  Argonne, a
				U.S. Department of Energy Office of Science laboratory, is operated
				under Contract No. DE-AC02-06CH11357.  The U.S. Government retains for
				itself, and others acting on its behalf, a paid-up nonexclusive,
				irrevocable worldwide license in said article to reproduce, prepare
				derivative works, distribute copies to the public, and perform
				publicly and display publicly, by or on behalf of the Government. The Department of Energy will provide public access to these results of federally sponsored research in accordance with the DOE Public Access Plan. http://energy.gov/downloads/doe-public-access-plan.
			}
		}
		\normalsize
	\end{center}
	
	\begin{IEEEbiographynophoto}{Tianyang Yi} (Student Member, IEEE)
		is currently
		working toward a Ph.D. degree in Industrial Engineering at the Pennsylvania State University. He is interested in developing practical methodologies for planning and optimization of modern power systems under uncertainty. 
	\end{IEEEbiographynophoto}
	
	\begin{IEEEbiographynophoto}{Daniel Adrian Maldonado} (Member, IEEE) received %
		the Ph.D. degree in electrical engineering from the Illinois Institute of Technology, Chicago, IL, USA,
		in 2017. He 
		is currently an Computational Scientist with
		Argonne National Laboratory, Lemont, IL, USA. His
		research interests include the application of high performance computing, uncertainty quantification,
		and statistical methods to power systems.
	\end{IEEEbiographynophoto}
	
	\begin{IEEEbiographynophoto}{Anirudh Subramanyam} %
		is currently an Assistant Professor in the Department of Industrial and Manufacturing Engineering at the Pennsylvania State University. He specializes in computational methods for optimization under uncertainty to address problems in power, energy, and infrastructure systems.
	\end{IEEEbiographynophoto}

\end{document}